\documentclass[preprints,article,accept,oneauthor]{Definitions/mdpi}
\usepackage{url}

\usepackage{enumitem} 
\setitemize{parsep=6pt,itemsep=0pt,leftmargin=*,labelsep=5.5mm}
\setenumerate{parsep=6pt,itemsep=0pt,leftmargin=*,labelsep=5.5mm}
\setlist[description]{itemsep=0mm}   
\usepackage{subcaption} 
\usepackage{wrapfig}
\usepackage{array}
\usepackage{comment}
\usepackage{url}
\usepackage{lineno}
\usepackage{tabularx}
\usepackage{booktabs}

\usepackage{braket}
\usepackage{siunitx}
\usepackage{revsymb}

\newcommand{\be}{\begin{equation}}
\newcommand{\ee}{\end{equation}}
\newcommand{\bea}{\begin{eqnarray}}
\newcommand{\eea}{\end{eqnarray}}

\firstpage{1} 
\pubvolume{00}
\issuenum{0}
\articlenumber{0}
\pubyear{2023}
\copyrightyear{2023}
\updates{yes}

\history{Arxiv manuscript submitted July, 2026}

\Title{Is The H Atom Surrounded By A Cloud of Virtual Quanta Due To The Lamb Shift?}

\Author{{G. Jordan Maclay \orcidA{}}}
\AuthorNames{G. Jordan Maclay}
\address{%
$^{1}$ \quad Quantum Fields LLC, Saint Charles, IL 60174 USA, jordanmaclay@quantumfields.com; \\
$^{2}$\quad Dept. of Electrical Engineering and Computer Science, University of Illinois at Chicago, Chicago, IL 60607, USA} 

\abstract{The Lamb shift, one of the most fundamental interactions in atomic physics, arises from the interaction of H atoms with the electromagnetic fluctuations of the quantum vacuum.  The energy shift has been computed in a variety of ways. The energy shift, as Feynman and Power demonstrated, equals the change in the vacuum energy in the volume containing the H atoms due to the change in the index of refraction arising from the presence of the H atoms. By using this result and a group theoretical calculation of the contribution to the Lamb shift from each frequency of the vacuum fluctuations, we can obtain an expression for the size of the region of vacuum energy for each frequency \texorpdfstring{$\omega$}{ω} around the H atom due to the Lamb shift.  The ground state atom is surrounded by a region of positive vacuum energy that extends well beyond the atom for low frequencies. This region can be described as a steady state cloud of virtual quanta. For energies \texorpdfstring{$E=\hbar\omega$}{E=hbar omega} eV less than 1 eV, the radius of the positive energy region is approximately 14. 4/E Angstroms. For a vacuum fluctuation of wavelength \texorpdfstring{$\lambda$}{lambda} the radius is \texorpdfstring{$(\alpha/2\pi)\lambda$}{(alpha/2pi) lambda}. Thus, for long wavelengths, the region has macroscopic dimensions.  The energy-time Uncertainty Relation predicts a maximum possible radius that is larger than this by a factor of \texorpdfstring{$1/ 4\alpha$}{1/(4 alpha)}.} 

\keyword{Bethe; radiative shift; shift spectral density; spectral volume; vacuum fluctuations; vacuum field; Lamb shift; QED; energy field, renormalization, zero point fluctuations; hydrogen atom}

\begin{document}

\section{Introduction}

Feynman called the three page long 1947 non-relativistic Lamb shift calculation by Hans Bethe the most important calculation in quantum electrodynamics because it tamed the infinities plaguing earlier attempts\cite{feynman1}. The seminal process in the calculation was to subtract the divergent energy shift for free electrons from the expression for the total energy shift, which reduced the divergence to a manageable logarithmic divergence.  Bethe believed the radiative shift was primarily a non-relativistic phenomena and used a frequency cut off corresponding to the the mass of the electron.  When the sum over all states was evaluated numerically, his renormalized result gave a finite prediction of the energy difference between the $2S_{1/2}$ and $2P_{1/2}$ levels of the hydrogen atom that agreed with experiment\cite{lam1}\cite{bethe}. There are numerous higher–order contributions to the Lamb shift~\cite{grotch,beyer,tang2013} that 
have been computed with very high precision. For example, to secure the greatest accuracy possible, the upper limit of integration for the Bethe logarithm—originally taken by Bethe as the electron rest mass—was increased to about $10^{42}$ times the electron mass, yielding 
a precision of 23 digits~\cite{goldman2000}. 

This paper, however, is not focused on achieving extreme numerical accuracy. Instead, our goal is to develop a spectral interpretation of the dominant lowest–order non‑relativistic radiative shift, which is the contribution Bethe originally calculated and which accounts for approximately $97\%$ of the total Lamb shift.
This shift can be interpreted as arising from virtual transitions of the H atom induced by the quantum fluctuations of the electromagnetic field. Since the vacuum field contains all frequencies, virtual transitions to all states, bound and scattering, are possible.  These short lived virtual transitions result in a slight shift in the average energy of the atom, the radiative Lamb shift \cite{gjmradiative}. The radiative shift for the ground state is positive since all virtual transitions are to higher energy levels. This same mechanism shifts every atomic energy level. We note that the Lamb shift can also be described as an interaction of the electron with its own radiation field, yielding the exact same results as if calculated with the vacuum field \cite{gjmradiative}\cite{mil}. Thus the results in this paper do not depend on the presence of vacuum fluctuations. For most conditions, H atoms are primarily in the ground state, which is why our focus in on the ground state radiative shift. 

Both the Lamb shift and van der Waals forces can be interpreted in terms of the interactions of atoms with the quantum fluctuations of the electromagnetic fields. When only one atom is present, the interaction results in the field around the atom corresponding to the Lamb shift. If multiple atoms are present, these clouds affect neighboring atoms; along with the zero-point field, this interaction leads to the van der Waals force.

Feynman provided an alternative non‑relativistic description of the origin of the Lamb shift, which forms the conceptual basis for our calculation of the vacuum energy cloud. He expressed the Lamb shift in terms of the change in the energy of vacuum fluctuations caused by the presence of H atoms altering the index of refraction in the region surrounding the atom~\cite{feyn}. The change in refractive index modifies the frequencies present in the vacuum field, thereby altering the energy. The detailed calculation was carried out by Power, who showed that the shift in the energy of the vacuum field around an H atom in a large box exactly equals the non‑relativistic radiative shift predicted by Bethe~\cite{power,maclayphysics,mil}. 
In Power’s treatment, the divergent contribution from free electrons must be 
subtracted to obtain the same finite result as Bethe. In this paper, we do not explore the details of Power’s derivation; instead, we use the conclusion that the change in vacuum‑field energy surrounding the atom equals the Lamb shift.

In a similar spirit, Milonni computed the shift in the atomic energy level due to the Stark effect arising from the atom’s interaction with the vacuum field, showing that the change in the atomic energy level exactly equals the change in the energy of the vacuum fluctuations, and that both equal the Lamb shift as computed by Bethe~\cite{milx} 
(p.~438).

We have previously calculated the non‑relativistic Lamb shift using SO(4,2) group 
theory to transform Bethe’s renormalized expression for the shift before any 
approximations were introduced to simplify its evaluation. The level shift is expressed as an integral over a shift spectral density as a function of the frequency of the vacuum fluctuations~\cite{maclayphysics,gjmdynam}. (In Ref.~\cite{gjmdynam}, two typographical errors appear: in Equation~(297) the right‑hand side should contain a plus sign rather than a minus sign; in Equation~(299) the integral should contain a minus sign and the $\ln$‑term should read $+\delta_{LO}\ln\!\left[2/(Z\alpha)^2\right]$.) 
Unlike Bethe’s original evaluation, no sum over states is required. This formulation provides an analytical expression for the contribution of each vacuum‑field frequency to the radiative Lamb shift, enabling us to compute the volume associated with the spectral components present in the Lamb shift.

The results in \cite{gjmradiative,maclayphysics,gjmdynam} and in this paper are all included in the recently published book \cite{hatombook}, along with additional information and corrections. A recent journal publication extends the SO(4,2) approach in this book to include the complex shift in energy levels (the level shift and width) and the recoil reaction \cite{alber}.

The calculations by Power and Milonni show that for the ground state 1S Lamb shift, which is positive, the vacuum field energy density around the atom must increase so that the integral of the energy over the volume surrounding the atom gives the 1S Lamb shift. The increased energy is supplied by the only source of energy available, the vacuum fluctuations. By comparing the needed vacuum energy obtained from our calculation of the spectral shift density \cite{maclayphysics,gjmdynam,hatombook} with the know energy density of the free vacuum fluctuations, it is possible to compute the volume of vacuum energy needed for each spectral component of the shift.  For energies above about 100 eV, the spectral volume is much smaller than the region occupied by the ground state wavefunction; for energies below about 1 eV, the spectral volume is significantly larger than the ground state wavefunction.  Consequently the focus of this paper is on the low energy regime. For this regime, we will show that the radius of the spherical spectral volume for a vacuum fluctuation of wavelength $\lambda$ is approximately $(\alpha/2 \pi)\lambda$, where $\alpha$ is the fine structure constant. A simple estimate of the size of the virtual photon cloud based on the uncertainty relation for energy and time predicts a maximum radius of the spectral volume which is larger than that predicted by the Lamb shift model by a factor of $1 / 4\alpha$ \cite{gjmradiative,hatombook}\cite{passante}.

\section{Radiative Shift and Spectral Density Calculations}

The group theoretical approach is based solely on the Schrodinger and Klein-Gordon equations of motion in the non-relativistic dipole approximation. We obtain the result \cite{gjmdynam,hatombook}
\be
\Delta E_{NL}=\frac{2\alpha}{3\pi (mc)^2} 
\int_0^{\hbar \omega_c} dE\,
\bra{NL}\, p_i\,\frac{H-E_N}{H-(E_N-E)-i\epsilon}\, p_i\,\ket{NL},
\ee
where $E=\hbar\omega$ is the energy of the vacuum field, $\hbar$ is the reduced Planck constant, $\omega$ is the radial frequency, $c$ is the speed of light, $m$ is the electron mass, and $p_i$ denotes the momentum operator in the $i$‑th direction. The Hamiltonian is $H = \frac{p^{2}}{2m} - \frac{Z\alpha\,\hbar c}{r}$,
where $p^{2}$ is the momentum‑squared operator, $r$ is the position operator for the electron, $Z$ is the atomic number of the nucleus, and the states $\ket{NL}$ are the hydrogen‑atom energy eigenstates with energies $E_N$. The cutoff frequency $\omega_c$ satisfies $\hbar\omega_c = mc^{2} = 511~\text{keV}$.

Equation~(1) represents the renormalized shift; the divergent contribution from the free electron has been subtracted. Inserting a complete set of states,
$1 = \sum_{n} \ket{n}\bra{n}, \text{where }     H\ket{n} = E_n\ket{n}$,
yields Bethe’s expression for the finite observable shift $\Delta E_{NL}$ for the state $\ket{NL}$:

\be
\Delta E_{NL} =\int_0^{ \hbar\omega_c} dE \frac{2\alpha}{3\pi (mc)^2}\sum_n |\textbf{p}_{nN}|^2 \frac{(E_n-E_N)}{E_n - E_N + E -i\epsilon} .
\ee
Bethe performed the integration over $E$, assuming that $mc^2>>|E_n-E_N| $, to obtain his an expression for the Lamb shift for an S state with principle quantum number N \cite{bethe} 
\be
\Delta E_{N}^{Bethe}=\frac{4\alpha}{3\pi}(\frac{1}{mc})^{2}\sum_{n}|\textbf{p}_{nN}|^{2}(E_n-E_N) ln[ {\frac{mc^2}{E_n-E_N}}].
\ee
To simplify the evaluation of the sum, he took the $ln$ term out of the summation and replaced $E_n$ by an average energy.  We can also express the radiative shift for the 1S ground state as an integral over the renormalized spectral shift density $d\Delta E_1/dE$ \cite{maclayphysics,hatombook}
 \be
\Delta E_1 = \int_0^{mc^2} \frac{d\Delta E_1}{dE}dE.
 \ee
This equation is a definition of the spectral shift density. If a level shift is expressed as an integral over energy, the integrand is the spectral shift density.  Since the shift $\Delta E_1$ has the units of eV, the spectral shift density has the units of eV/eV or is dimensionless.  The spectral shift density from Bethe's formulation equals the integrand in Equation 2, which requires the evaluation of a sum over all states, including scattering states. A more convenient expression for the spectral shift density for the renormalized ground state can be obtained by transforming Eq.1 using group theory \cite{gjmdynam,hatombook} 
\be
\frac{d\Delta E_1}{dE}=\frac{4 \alpha^{3}}{3 \pi}  {e}^{-2\phi} \sinh \phi \int_{0}^{\infty} {ds e^{s{e^{-\phi}}}}\frac{1} {\sinh(\frac{s}{2})^{2}} \frac{1}{\left(\coth \frac{s}{2}+\cosh \phi \right)^{3}}.
\ee
The integral over $s$ can be evaluated exactly by Mathematica for specific values of the dimensionless normalized frequency variable $\phi$
\be
\phi=\frac{1}{2} ln[1+\frac{E}{|E_1|}]
\ee
where $E_1$ is the ground state energy -13.6 eV. The cutoff $\phi_c$ corresponds to $E=\hbar\omega_c=m c^2= 511$ keV corresponding to the electron mass.

Fig. 1 shows a loglog plot of the spectral density $\frac{d\Delta E_1}{dE} $ of the ground state Lamb shift with Z=1 over the entire range of energy $E$ computed from Eq. 5. For energies above about 100 eV, the spectral density is approximately proportional to 1/E, whereas below about 10 eV, the spectral density increases slowly to a maximum at E=0, as shown in Fig. 2.  

 \begin{figure}[htbp]
    \centering
    \includegraphics[width=0.9\linewidth, height=5cm]{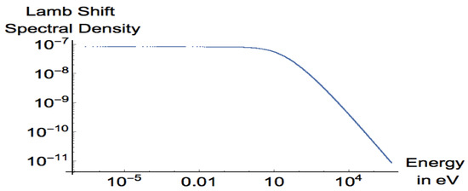}
    \caption{This loglog plot shows the log of the spectral density of the ground state shift from the group theoretical expression Eq. 5 on the vertical axis versus the log of the energy in eV. For energies above about 100 eV, the behavior is dominated by a 1/energy dependence. From about 10 eV to 0 eV, there is a slight linear increase in the spectral density. }
    \label{Fig. 1}
\end{figure}

\begin{figure}[htbp]
\includegraphics[width=0.95\linewidth, height=4.0cm]{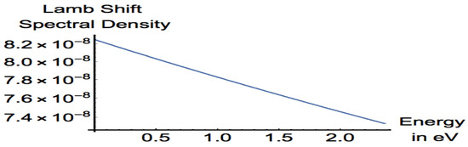}
\caption{Linear plot of the ground state spectral density as a function of eV calculated from group theory, plotted as a function of energy from 3 eV to 0 eV showing an approximately linear increase to its maximum value at 0 eV.}
\label{fig2}
\end{figure} 

The low energy limit of the group theoretical result Eq. 5 for the S state shift density can be taken analytically, giving \cite{maclayphysics,hatombook}
\be
\frac{d\Delta E_{n}}{dE}|_{E->0}=
\frac{2\alpha}{3\pi}\frac{(Z\alpha)^2}{n^2}-\frac{\alpha}{\pi mc^2}E
\ee
where n is the principle quantum number.
 The corresponding spectral density for the ground state with $n=1, Z=1$ is
\be
\frac{d\Delta E_{1}}{dE}|_{E->0}
=\frac{4\alpha \times 13.6}{3\pi mc^2}(1-\frac{3E}{4\times 13.6})= 8.253 \times 10^{-8}(1 - 0.0551E) .
\ee
which is in agreement with Figure 2.  As E decreases to zero the spectral density increases linearly to a constant value $\frac{4\alpha}{3\pi}\frac{|E_n|}{mc^2}=2\alpha^3Z^2/3\pi n^2 =8.253\times 10^{-8}/n^{2}$. The intercept goes as $1/n^2$, and the slope equals $\alpha/\pi m c^2$.   For $ E<1$ eV, the ground state spectral density equals $8.25 X 10^{-8}$ to within about $ 5\%$. 
\section{Computing the Size of Vacuum Energy Field}

 Consider a large box containing H atoms in the ground state.  The spectral density $\frac{d\Delta E_1}{dE} $ of the ground state shift and the energy density of the quantum vacuum with no H atoms present are both know.  In the box containing the H atoms the vacuum field density must increase so that the integral of the energy density over the volume gives the 1S Lamb shift.  This increase in vacuum energy is provided by the vacuum fluctuations, which have a free field energy density equal to \cite{mil}
\be
\rho_0(\omega)=\frac{\hbar\omega^3}{2\pi^2c^3}
\ee
where $c$ is in cm/s, $\omega$ is the radial frequency in s$^{-1}$, and $\rho$ has units of erg/cm$^{3}$–s$^{-1}$. If the frequency is measured in eV, then it is essentially the energy $E=\hbar\omega$, and the vacuum spectral energy density has units of $\frac{\text{eV}}{\text{cm}^{3}\,\text{eV}} = \frac{1}{\text{cm}^{3}}$,
and is

\be
\rho_0(E)=\frac{E^3}{2\pi^2 \hbar^3 c^3}  
\ee
so that the integral $\int_{E_1}^{E_2}\rho_0(E)dE$ would be the energy density eV/cc in the energy interval $E_1$ to $E_2$. The question being addressed is: what volume of vacuum energy of density $\rho_0(E)$ is required to supply the amount of energy corresponding to the radiative shift? The total renormalized radiative shift $\Delta E_1$ can be expressed as the integral of the vacuum energy density $\rho_0(E)$ over an effective volume $V_1(E)$ 
\be
\Delta E_1=\int_0^{mc^2} dE \rho_0(E) V_1(E)
\ee
with the same upper limit for $E$ as previously used \cite{maclayphysics,hatombook}.  Recall the definition of the spectral shift density Equation 4:
\be
\Delta E_1=\int_0^{mc^2} dE \frac{\Delta E_1}{dE}.
\ee
Comparison of Equation 11 and Equation 12 shows that to insure energy balance at each energy E, the effective spectral volume $V_1(E)$ is\cite{maclayphysics,hatombook}
\be
V_1(E)=\frac{d\Delta E_1}{dE}\frac{1}{\rho_0(E)}.
\label{eq13}
\ee
The spectral volume $V_1(E)$ has the dimensions of $cm^3$ and contains the amount of vacuum energy at energy value $E$ that corresponds to the ground state spectral density at the same energy $E$. 
Equations 11, 12 and 13 are general equations and apply to any calculation of the radiative Lamb shift that can be expressed as an integral over the vacuum energy, as in Equation 4. The utility of the Equations 11 - 13 lies in our ability to give an explicit analytical expression for the spectral shift using our group theoretical results [14]\cite{hatombook}. 

An example of Equation 11 appears in Milonni’s calculation of the Lamb shift as a 
Stark shift~\cite{mil}. Consider the energy 
\[
W = -\frac{1}{2}\,\mathbf{d}\cdot\mathbf{E}(\omega)
\]
for a dipole $\mathbf{d}$ in an isotropic field $\mathbf{E}(\omega)$. If the dipole is induced by the 
field, then $\mathbf{d}(\omega)=\alpha(\omega)\mathbf{E}(\omega)$, where $\alpha(\omega)$ is the 
polarizability. The energy for an atom $A$ at position $\mathbf{x}_A$ with polarizability 
$\alpha_A(\omega)$ can be written as~\cite{mil}
\begin{equation}
W_A = -\frac{1}{2}\int_{0}^{\infty} d\omega\, \alpha_A(\omega)\,\bra{}E^2(\omega)\ket{} .
\label{eq14}
\end{equation}
For the Lamb shift, the field correlation function satisfies 
$\bra{} E^2(\omega)\Ket{} = 4\pi\,\rho_0(\omega)$, where $\rho_0(\omega)$ is the zero-point 
vacuum spectral energy density. Substituting this into Equation~(\ref{eq14}) gives
\begin{equation}
W_A = -2\pi\int_{0}^{\infty} d\omega\, \alpha_A(\omega)\,\rho_0(\omega).
\label{eq15}
\end{equation}
The polarizability $\alpha_A(\omega)$ is provided by the Kramers--Heisenberg formula and has 
units of volume. Equation~(\ref{eq15}) has the same mathematical structure as 
Eq. 11. To complete the Stark-shift calculation, the contribution from free 
electrons must be subtracted; after this subtraction, the final result is identical to Bethe’s 
non-relativistic Lamb shift~\cite{mil}.

The spectral volume in Equation 13 is assumed to be spherical since we are dealing with S states, so the radius can be calculated from the known spectral volume. This assumption will be discussed more in Section 4. Fig. 3 shows a loglog plot of the radius of the spectral volume $V_{1}(E)$ for the ground state as a function of $E$ for the entire energy range. 
 Fig. 4 shows the radius in Angstroms of the spherical spectral volume for energies below 23 eV.  For an energy of 21.7 eV, the spectral radius equals the mean radius of the ground state wavefunction 0.53 Angstroms. For energies less than 21.7 eV, the radius will be greater than the ground state radius.  For an energy of 1 eV, the radius is 14 A.  This calculation predicts that there is a sphere of positive vacuum energy of radius 14 A around the atom corresponding to the 1 eV shift spectral density.

\begin{figure}[ht]
    \centering
    \includegraphics[width=0.94\textwidth,height=5cm]{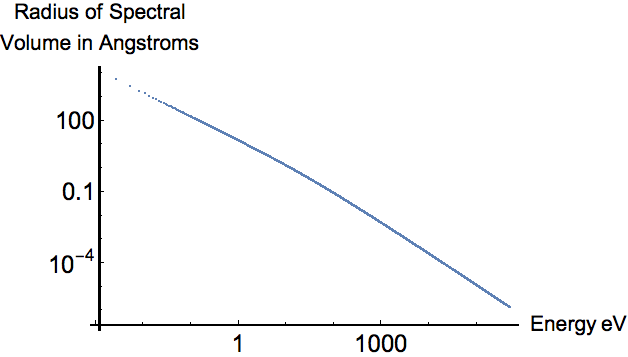}
    \caption{The log of the radius in Angstroms of the spherical spectral volume $V_1(E)$ as a function of the log of the vacuum field energy E from 0.0027 eV, where the radius is 5330 Angstroms, to 511,000 eV, where the radius is $10^{-17}$ Angstroms.}
\end{figure}

\begin{figure}[ht]
    \centering
    \includegraphics[width=0.94\textwidth,height=5cm]{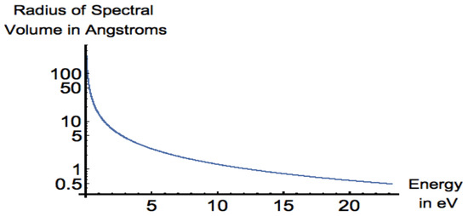}
    \caption{This plot shows the log of the radius in Angstroms of the spherical spectral volume $V_1(E)$ as a function of the vacuum field energy E from 0.05 eV to 23 eV, with corresponding radii of 288 Angstroms and 0.5 Angstroms.}
\end{figure}

For low energy virtual photons, from Equation 7 the spectral density for an S state with principle quantum number $n$ can be approximated as a constant
\be
\frac{d\Delta E_{n}}{dE}|_{E->0}=
\frac{2\alpha}{3\pi}\frac{(Z\alpha)^2}{n^2}.
\label{eq16}
\ee
This is accurate to about 5\% at 1 eV and the accuracy increases as the energy decreases. This approximation corresponds to the end point $E=0$ of the nearly horizontal portion of the spectral density in Fig. 1. For these low energies the spectral volume $V_n(E)$ from Eq.13 is
\be
V_n(E)=\frac{4\pi}{3} \frac{\alpha (Z\alpha)^2}{n^2} \frac{\hbar^3 c^3}{E^3}.
\ee
Assuming a spherical spectral volume of radius $R_V(E)$, we obtain for a state $n$
\be
R_V(E)=\left[ \frac{\alpha(Z\alpha)^2}{n^2} \right]^{1/3} \frac{\hbar c}{E}
\ee
which gives for the 1S state of hydrogen
\be
R_V(E) = \alpha\,\frac{\hbar c}{E} = \frac{14.4\,\AA}{E\ \text{(eV)}} \mathring{\text{A}}
\label{eq19}
\ee

The H atom is surrounded by a steady-state cloud of radius $R_V(E) $ of virtual quanta continuously emitted and reabsorbed by the field. 
These virtual quanta cannot leave the vicinity of the atom since they are reabsorbed.  This vacuum energy density of the cloud is positive in the sense that it is above the free field vacuum energy density. 

It is remarkable that the asymptotic low energy spectral radius for the ground state $R_V(E) $ has such a simple form, Equation 19. This result can also be rewritten using the definition $\alpha=e^2/\hbar c$ as
\be
E= \frac{e^2}{R_V(E)}.
\label{eq20}
\ee
Thus the Coulomb energy for two electrons a distance $R_V(E)$ apart equals the energy $E=\hbar \omega$ of the corresponding vacuum virtual photon.

It is interesting to compare the radius $R_V(E)$ of the spectral volume  with the wavelength $\lambda$ of the vacuum fluctuation corresponding to $E=\hbar\omega=2\pi\hbar c/\lambda$ which gives for the ground state
\be
R_V(E)=\frac{\alpha}{2\pi} \lambda=\frac{\lambda}{861}\mathring{\text{A}}.
\ee
The radius of the spectral volume equals $\alpha/2\pi$ times the wavelength of the corresponding vacuum fluctuation. Thus long wavelength vacuum fluctuations produce macroscopic regions of positive vacuum energy for the hydrogen ground state.  
\subsection{Comparison to Predictions from the Uncertainty Relation}
A simple analysis using the uncertainty relation can provide an order of magnitude estimate of the largest extent of the positive energy vacuum field. The hydrogen atom is a quantum system and consequently its ground state energy can vary for a time interval $\tau$ by an amount $\Delta E_u$ which is constrained by the uncertainty relation \cite{milx}(p. 201)
\be
\Delta E_u \tau < \hbar/2.
\ee
The variation in energy is due to the emission and absorption of virtual photons of energy $\Delta E_u =\hbar \omega_u$ and frequency $\omega_u$. Since the velocity of the photon is $c$, in the time $\tau$ it can travel a distance $2R_u$ where \cite{passante}
\be
R_u< \frac{\hbar c}{4\Delta E_u} = \frac{c}{4\omega_u}=\frac{\lambda}{8\pi}
\ee
where $\lambda$ is the wavelength of the virtual photon. Comparing Equation 23 to Equation 21 shows that for the same energy virtual photon, 
\be
R_V=4\alpha R_u.
\ee
For virtual fluctuations of energy E below about 1 eV, the dimension $R_V$ of the virtual cloud predicted by an analysis of the ground state Lamb shift is $4\alpha$ times smaller than the maximum extent $R_u$ allowed by the uncertainty relation. Ref.~\cite{passante} has suggested that ${(4\pi/3)}\,\alpha$ can be interpreted as the mean density of virtual photons in the region surrounding the atom, which may explain the difference between $R_{u}$ and $R_{V}$.

\section{Significance of the Zero-Point Field around the~Atom}\label{sec4}

The cloud of quantum fluctuations surrounding the H atom can be interpreted as
resulting from the scattering of the free-field vacuum fluctuations by the atom.
The~zero-point field activates the atom in a continuous process, creating the steady-
state cloud of quantum fluctuations that is 
described in this paper.
As~the derivation of the Lamb shift in terms of the Stark effect suggests, the~zero-point
field induces an instantaneous dipole moment in the atom
that leads to a dipole field. The continuous~stochastic excitation from the zero-point
field leads to a sum of incoherent contributions that
average to a spherically symmetric cloud~ {\cite{milx}} .

{One} 
can imagine the atom undergoing virtual transitions from the ground state to all higher
energy states and~then returning to the ground state in~accordance with the time--
energy uncertainty relation. For~a zero-point fluctuation of the wavelength $\lambda$,
{our} calculations show
that the cloud extends about $\alpha\lambda/2\pi$ from the nucleus, which can be a
macroscopic distance. Thus far, direct measurement of such vacuum fluctuations has
eluded experimentalists. There are two ways
{of} 
exploring the significance of this cloud of vacuum energy: first, by computing estimates
of the mean energy density; and second, by explaining its role in the creation of van der
Waals forces under the assumption that another H atom is~nearby.

\subsection{Energy Density of the Zero-Point Field around the~Atom}Using
the results
In Section 3, it is possible to compute the energy density of this field as a function of
distance for different wavelengths or energy
intervals of the zero-point field. Consider a spherical shell: the~inner radius corresponds
to one energy and is given by
Equation~ \eqref{eq19}.
The outer radius corresponds to a slightly smaller energy. For~this energy interval,
{one} can estimate the contribution to the total Lamb shift by integrating the curve in
Figure~\ref{Fig. 1}~\cite{maclayphysics,hatombook}.
For~energies below 1 eV the~contribution to the ground state Lamb shift is about
0.24\% of the total shift. In~this low energy {range,}
the~contribution to the shift scales linearly with the energy, as shown in
Figure~\ref{fig2}. This allows us to compute a mean energy
{density, $\rho_{\rm LS}^{\rm shell}$,} of the quantum fluctuations in a spherical~shell.

\begin{figure}[H]
\includegraphics[width=0.94\textwidth,height=5cm]{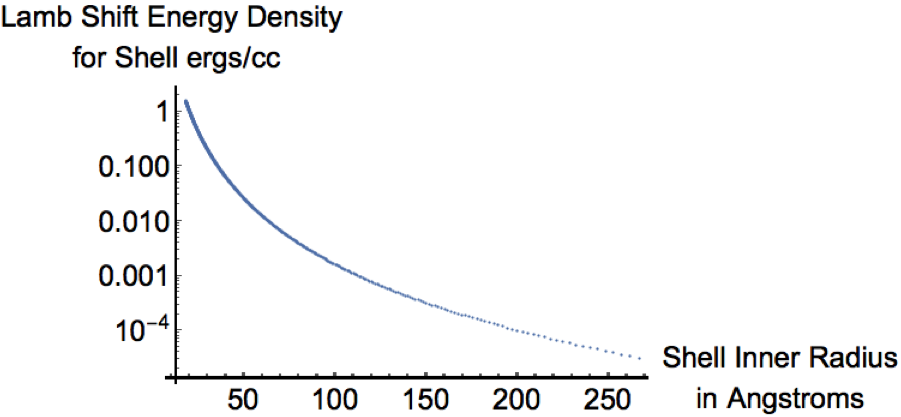}

\caption{The Lamb shift energy {density, $\rho_{\rm LS}^{\rm shell}$} Equation
{(\ref{eq27})} 
as a function of the inner {radius, $R_1=\alpha
{\hbar c}/{(\beta E)}$,} 
of
the shell. 
The~outer {radius, $R=\alpha {\hbar c}/{E}$,} 
is $\beta=1.03$ times the inner radius; thus, $g(\beta)=11.3$
{(see} Equation (\ref{eq29})).
}
\label{fig:my_label5}
\end{figure}

The density of the Lamb shift energy in the spherically symmetric
region of vacuum energy surrounding the H atom can be analyzed in terms of shells
with an outer radius
of
{$R=\alpha {\hbar c}/{E}$} 
and inner radius of
{$R_1=\alpha{\hbar c}/{E_1}$.}
It is convenient to let $E_1=\beta E$, where $\beta > 1$. Assuming that both energies
are less than 1 eV,
{one} 
{ can integrate} the Lamb shift
spectral density from Equation~ \eqref{eq16} 
for the ground state and {$Z$} = 1 to~obtain

\be
\Delta E_{\rm LS}^{\rm shell}(E)
    = \int_E^{\beta E} dE \frac{2\alpha^3}{3\pi}
    = \frac{2\alpha^3}{3\pi} E(\beta - 1)
\ee

to an accuracy of about 5\%.
The volume of the shell is
\be
V^{\rm shell}(E)= \frac{4\pi}{3}(\alpha\hbar c)^3\left({1-
\frac{1}{\beta^{3}}}\right)\frac{1}{E^3}.
\ee
{Therefore} the Lamb shift energy density 
{(in erg/cm$^3$)}
in the shell {is} 
\be
\rho_{\rm LS}^{\rm shell}(E)=\frac{\Delta E_{\rm LS}^{\rm shell}(E)}{V^{\rm
shell}(E)}=\frac{1}{2 \pi^2}\frac{1}{(\hbar c)^3}\frac{\beta
-1}{1-1/\beta^3}E^4.
\label{eq27}
\ee
{The} Lamb shift energy density for a shell with outer {radius
$R=\alpha \hbar c/E$,} and inner {radius $R_1=\alpha \hbar c /(\beta E)$,}
is 
proportional to $E^4$ or $1/R_V^4(E)$. Figure~\ref{fig:my_label5} shows the value of$\rho_{\rm LS}^{\rm shell}$ as a function of the inner
radius {(in \AA)},
where the outer radius is 1.03 times the inner radius ($\beta=1.03$).

One can compare {the}
energy density $\rho_{\rm LS}^{\rm shell}(E)$ from Equation {(\ref{eq27})} {(in
erg/cm$^3$)}
to the energy density $\rho_0^{\rm shell}(E)$
{(in erg/cm$^3$)}
of the free zero-point
vacuum field for the same spectral interval, i.e., from~$E$ to $\beta E$:
\be
\rho_0^{\rm shell}(E)=\int_E^{\beta E} dE \rho_0(E)= \frac{1}{8 \pi^2
\hbar^3c^3}E^4(\beta ^4-1).
\label{eq28}
\ee
{One finds that the ratio} 
\be
\frac{\rho_{\rm LS}^{\rm shell}(E)}{\rho_0^{\rm shell}(E)}=4\frac{\beta -1}{1-
{1}/{\beta^3}}
\frac{1}{\beta^4-1}=g(\beta),
\label{eq29}
\ee
is a constant that depends on $\beta$.
The Lamb shift energy density for the shell is directly proportional to the free vacuum
energy density for the same energy
interval. This result follows for low {$E$} since the spectral {density, $\frac{d\Delta
E_1}{dE}$} from
Equation~ {(\ref{eq16})}, 
is a constant. Comparison with
Equation~ {(\ref{eq13})} 
shows that $\rho_0(E)V_1(E)$ is therefore constant and independent of {$E$} for low~
{$E$} {values}.

The function $g(\beta)$ is singular at $\beta=1$ and decreases rapidly as $\beta$
increases. For~$1 < \beta <1.35$, $g(\beta)$ is greater
than 1. For~$\beta$ of (1.01, 1.02 1.03, 1.05, 1.1), the~corresponding values of
$g(\beta)$ are $(33.5, 16.8, 11.3, 6.8, 3.5)$. For~these
shells, $\rho_{\rm LS}^{\rm shell}(E)$ is always greater than $\rho_0^{\rm shell}(E)$.
Just as free-field vacuum fluctuations are important in
many physical systems, the field of fluctuations due to the Lamb shift must be
equally~important.

Table 1 shows the results of computing the energy densities for different
spherical {shells.} The~first row corresponds to a shell with the frequency range of the visible spectrum (400 nm to 700 {nm),} for~which the energy
density in the {shell, 
$\rho_{\rm LS}^{\rm shell}$,} is {98.7} {erg/cm$^3$}, 
which is 45\% of the corresponding $\rho_{0}^{\rm shell}$ (the~free-field energy density for the shell) of~{218} {erg/cm$^3$}. 
For~cases with $\beta < 1.35$, 
the~ratio $\rho_{\rm LS}^{\rm shell}/ \rho_0^{\rm shell}$ in the fifth column is greater than~one.

\begin{table}[H]
\caption{The inner and outer radii for a spherical shell around the atom, the corresponding fluctuation energies $\rho_{\rm shell}^{\rm LS}$ and $\rho_{\rm shell}^{0}$, and the ratio $\rho_{\rm shell}^{\rm LS}/\rho_{\rm shell}^{0}$.}

\label{tab:shelltable}

\newcolumntype{C}{>{\centering\arraybackslash}X}

\begin{tabularx}{\textwidth}{p{3.8cm} p{3.8cm} p{3.2cm} p{2.5cm} p{2.5cm}}
\toprule
\textbf{Inner and Outer Radii (Å)} &
\textbf{Energy Range (eV)} &
\textbf{$\rho_{\rm shell}^{\rm LS}$ (erg/cm$^3$)} &
\textbf{$\rho_{\rm shell}^{0}$ (erg/cm$^3$)} &
\textbf{$\rho_{\rm shell}^{\rm LS}/\rho_{\rm shell}^{0}$} \\
\midrule

4.64–8.13     & 3.10–1.77       & 98.7               & 218                 & 0.45 \\
10–20         & 0.72–1.44       & 3.34               & 10.65               & 0.314 \\
20–30         & 0.48–0.72       & 0.399              & 0.570               & 0.700 \\
30–40         & 0.36–0.48       & 0.1024             & 0.0959              & 1.068 \\
40–50         & 0.288–0.36      & 0.0373             & 0.0262              & 1.16 \\
50–60         & 0.240–0.288     & 0.0167             & 0.00941             & 1.77 \\
60–70         & 0.2057–0.240    & 0.00852            & 0.00403             & 2.12 \\
70–80         & 0.180–0.2057    & 0.00480            & 0.00196             & 2.45 \\
80–90         & 0.160–0.180     & 0.00291            & 0.00104             & 2.79 \\
90–100        & 0.144–0.160     & 0.00186            & 0.000595            & 3.13 \\
140–150       & 0.096–0.1029    & 0.000344           & 0.0000718           & 4.78 \\
200–210       & 0.0686–0.072    & 8.59$\times10^{-5}$ & 1.25$\times10^{-5}$ & 6.87 \\
300–310       & 0.0465–0.048    & 1.76$\times10^{-5}$ & 1.67$\times10^{-6}$ & 10.49 \\
400–410       & 0.0351–0.036    & 5.63$\times10^{-6}$ & 4.27$\times10^{-7}$ & 13.16 \\
1000–1020     & 0.01412–0.0144  & 1.46$\times10^{-7}$ & 8.58$\times10^{-9}$ & 16.97 \\
\bottomrule
\end{tabularx}
\end{table}

The energy densities $\rho_{\rm LS}^{\rm shell}$
for the shells are significant, for~example, compared to the energy densities,
$\rho_{\rm bb}$,
for black body radiation over the same spectral intervals. For~a temperature of 600 K
(for which the peak intensity is at about 5 micrometers or 0.25 eV) the ratio of
$\rho_{\rm LS}^{\rm shell}/\rho_{\rm bb}$ is 2.8
{$\times$} $10^{4}, 148$, and~$10.1$, respectively, for the shells with radii 20 {--
}
30 {\AA}, 
50--60 {\AA}, 
and~200--210 {\AA}. 
{Of course} black body radiation is ordinary electromagnetic radiation, while the Lamb shift energy consists of vacuum fluctuations of the electromagnetic~field.

\subsection{Relationship between the Zero-Point Field around the Atom and van der
Waals~Forces}
Zero-temperature Lamb shifts and van der Waals interactions have
{straightforward}
physical interpretations in terms of fluctuating zero-point fields~\cite{mil, milonni2023}. Here, we have considered
an isolated atom A and~described the fluctuating field around this atom that arises from
its interaction with the free field vacuum fluctuations. The~field around atom {A}
corresponds to the non-relativistic Lamb shift for atom {A}. If~another atom is present,
the~field around {A} plays an essential role in the van der Waals forces between
the~atoms.

To illustrate this, 
generalize
Equation~ {(\ref{eq14})} 
for the energy of an induced dipole at {A} to include a second atom {B.} %
The~total field is $\textbf{E}_{\textbf{k},\omega}$ and the combined energy is
\cite{mil}(Section 3.11)

\be
W_{AB}=-
\frac{1}{2}\sum_{\textbf{k}\omega}\alpha_A(\omega_k)\bra{}\textbf{E}_{\textbf{k}\omega}2(\textbf{x}_A,t)\ket{},
\ee
where $\alpha_A(\omega_k)$ is the polarizability of atom {A}, {$\textbf{k}$ is} the
wave vector, {and $t$} denotes the time.
The~presence of the second atom breaks the spherical symmetry, so
a summation {over}
$\textbf{k}$ for the non-isotropic field is included. The~total field acting on {A} is
assumed to be the sum of the zero-point
{field $\textbf{E}_{0,\textbf{k}\omega}(\textbf{x}_A,t)$} acting on {A} and the field
produced at {A} by atom {B}:
\be
\textbf{E}_{\textbf{k}\omega}(\textbf{x}_A,t)=\textbf{E}_{0,\textbf{k}\omega}(\textbf{x}_A,t
)
+\textbf{E}_{B,\textbf{k}\omega}(\textbf{x}_A,t).
\ee
{Each} atom is "driven" by the zero-point field at its location, creating a fluctuating
dipole field about the atom. The~field about atom {B} affects atom {A}
and~vice~versa.
The portion of the 
energy $W_{AB}$ that depends on the distance between the atoms corresponds to the
van der Waals force, and {is} 
\label{eq32}
\be
W_{AB}^{\rm vdW}
=-
\frac{1}{2}\sum_{\textbf{k}\omega}\alpha_A(\omega_k)\bra{}\textbf{E}_{0,\textbf{k}\omega}(\textbf{x}_A,t)\textbf{E}_{B,\textbf{k}\omega}(\textbf{x}_A,t)+\textbf{E}_{B,\textbf{k}\omega}(\textbf{x}_A,t)\textbf{E}_{0,\textbf{k}\omega}(\textbf{x}_A,t)\ket{}.
\ee
The term $\alpha_A(\omega_k)\bra{}\textbf{E}_{0,\textbf{k}\omega}
(\textbf{x}_A,t)\ket{}^2$ in the summation in Equation 30 does not depend on the
separation between the atoms, and corresponds to the Lamb shift for atom {A}. This is
the field of vacuum fluctuations about the atom that represents the atoms response to
the vacuum field described From~Equation~ {({32})}, 
{one can immediately see}
that this field plays an essential role in the van der Waals force. Similarly, this~field
would be essential for the Casimir--Polder force between an atom and a
surface~\cite{mil}.

The term $\textbf{E}_{B,\textbf{k}\omega}(\textbf{x}_A,t)$
represents the field from the induced dipole at atom {B}, and~is proportional to the
{polarizability $\alpha_B(\omega_k)$} of atom {B}. After computation, the~final
expression for the van der Waals force is shown
to be a
symmetric integral over $\omega$ of the
product $\alpha_A(\omega) \alpha_B(\omega)$ times a function of $\omega$ and
$r=|\textbf{x}_A-\textbf{x}_B|$. 

Along with the free field, the field about the atom corresponding to the Lamb shift affects
the nearby atom and results in a fluctuating induced dipole moment. The~correlation
between the fluctuating dipole moments at the two locations gives rise to the van der
Waals forces. The~correlation falls off rapidly with frequency and with the distance $r$
between the two locations, giving the $r^{-6}$-dependence of the non-retarded van der
Waals interaction~\cite{mil}. The~cloud of zero-point fluctuations about the H
atom
{described}
in this paper is fundamental to van der Waals forces as well as to the Lamb shift.
These~phenomena are linked in that they both arise from the interaction between atoms
and the fluctuating zero-point~field.

The van der Waals forces tend to become retarded for distances greater than about
$a_0/\alpha$ ($a_0$ being the Bohr radius of the ground-state wavefunction), or about
$70$ {\AA}. 
From~the calculations in Table 1, one can see
that lower energy fluctuations are responsible for these dispersion~forces.

\section{Conclusion}

The non-relativistic Lamb shift can be interpreted as due to the interaction of an atom with the fluctuating electromagnetic field of the quantum vacuum. The renormalized radiative Lamb shift can be expressed in terms of a spectral shift density which is a function of frequency $\omega$ or the energy $E=\hbar \omega$ of the vacuum field. The integral of the spectral density from E=0 to the rest mass energy of an electron, 511 keV, gives the non relativistic radiative shift for that state of the atom. Feynman, Power, and Milonni showed that the radiative shift also equals the change in the energy of the vacuum fluctuations in a region containing the H atom.  By using this result and a group theoretical calculation of the contribution to the Lamb shift from each frequency of the vacuum fluctuations, we can obtain an expression for the size of the region of vacuum energy for each vacuum energy E around the H atom due to the Lamb shift. The spectral volume for energy E around a H atom contains vacuum fluctuations of energy E and the total energy of these fluctuations equals the radiative shift corresponding to that energy E.  For the ground state, the energy density in the spectral volume is positive, meaning it is above the free field energy density. For E>23 eV, the radius of the region of positive energy vacuum fluctuations is less that the atomic radius. On the other hand, for energies less than 1 eV, the radius is approximately $\alpha \hbar c/E=14.4/E$ Angstroms, and can be much larger than the ground state wavefunction.  The radius of the spectral volume can also be expressed in terms of the wavelength of the corresponding vacuum fluctuations as $\alpha \lambda/2 \pi=\lambda/861$.  A simple estimate of the extent of photons from virtual transitions based on the Uncertainty Relation for time and energy predicts a maximum radius that is about $1/4\alpha$ larger that the radius based on the radiative shift calculations.  

The vacuum energy field around the H atom described in this paper plays an essential role in the van der Waals forces as well as in the Lamb shift. These phenomena are linked since both arise from the interaction between atoms and the fluctuating zero-point field.

The calculations in this paper were performed for the ground state of H, which has
a positive radiative shift. States with a negative radiative shift, such as 2P, would have a spectral volume as well; however, the energy would be negative, i.e., below the free-field vacuum energy. Notably, this analysis is complicated by the fact that the 2P state decays to the 1S state.

\section*{Funding}
This research received no external funding. The author has no conflicts of interest.
\section*{Acknowledgements} I thank Prof. Peter Milonni for many insightful and enjoyable discussions and for his helpful comments on this paper.

\end{document}